# Investigating charge transfer dynamics at the nanoscale

Hatem Labidi[1], Henry Pinto[2], Jerzy Leszczynski[2] and Damien Riedel[1*]

[1] *Institut des Sciences Moléculaires d'Orsay (ISMO), CNRS, Univ Paris Sud, Université Paris-Saclay, F-91405 Orsay, France*

[2] *Interdisciplinary Center for Nanotoxicity, Department of Chemistry, Jackson State University, Jackson, Mississippi 39217, USA*

**Abstract:**

*Acquiring quantitative information on charge transfer (CT) dynamics at the nanoscale remains an important scientific challenge. In particular, CT processes in single molecules at surfaces needs to be investigated to be properly controlled in various devices. To address this issue, the dynamics of switching molecules can be exploited. Here, a Nickel-tetraphenylporphyrin adsorbed on the Si(100) surface is used to study the CT process ruling the reversible activation of two chiral molecular conformations. Via the electrons of a scanning tunneling microscope (STM), a statistical study of the molecular switching reveals two specific locations of the molecule for which their efficiency is optimized. The CT mechanism is shown to propagate from two lateral aryls groups towards the porphyrin macrocycle inducing an intramolecular movement of two symmetric pyrroles. The measured switching efficiencies can thus be related with a Markus-Jordner model to estimate relevant parameters that describe the dynamics of the CT process. Numerical simulations provide a precise description of the molecular conformations and unveil the molecular energy levels that are involved in the CT process. This quantitative method opens a completely original approach to study CT at the nanoscale.*

\* Corresponding author: damien.riedel@u-psud.fr

**INTRODCUTION.**

Charge transfer (CT) processes are at the heart of our daily life as they rule general phenomena such as the ones observed in electronic transport[1], solar cell[2], organic emitting devices[3], spintronics[4] or biological systems[5]. Whether CT involve photoinduced excited states or ionic species, the perturbation and the scale at which the CT processes are described is of crucial importance. Heretofore, CT processes have been intensively studied in gas phase or in solution and the precise time scale involved in their dynamics can only be reached via complex tools such as pump-probe experiments using short pulsed lasers[6]. However, at the molecular scale, these techniques cannot be easily tailored, in particular with single molecule on surfaces. While it is clear that the investigation of charge transfer processes at the nanoscale is attracting an increasing interest, such studies remains hardly accessible. To circumvent this problem, it is necessary to define new methods with a precise and controlled environment of the studied molecular device at the nanoscale[7]. For this purpose, scanning probe microscopy is certainly the best available experimental techniques as it can observe and acts on molecules at the atomic scale via optically[8,9] or electronically[10] induced excitations. This has resulted in tremendous progresses in discovering new properties of individual molecular devices by controlling various effects such as the surface isomerization[11], charge storage[12], light molecular emitters[13] or molecular magnets[14]. While the activation of a basic molecular reaction on a surface was a significant step[15], one of the most challenging topics in this domain has been to develop controllable molecular switches[16] on surfaces[17] because it represents basic building blocks for the design of molecular architectures and machines[18]. Moreover, molecular switches provide the ability to study a large variety of reversible states offering a set of significant data at the atomic scale in relation to their molecular dynamics[7,19,20]. In this context, the molecule-surface interactions is shown to be of relevant importance on the molecular dynamics[11].Yet, STM techniques cannot resolve directly the time scale of the molecular dynamics, though several related processes could be partly described by combining STM excitations and numerical simulations with density functional theory (DFT)[21,22,23]. Hence, using molecular switches to probe the dynamics of CT at the nanoscale is a promising solution that has, so far, never been realized.

In this article, we use a single 5,10,15,20-Nickel(II)-Tetraphenyl-porphin (NiTPP) molecule adsorbed on the bare Si(100)-2x1 that can be activated with the electrons of a STM tip running at low temperature (9 K) to form an intra-molecular switch. By measuring quantitatively the intramolecular switching yield as a function of the excitation energy at different positions, we show that the electronically induced excitation is the starting point of a CT process which dynamics rules the switching efficiencies. DFT simulations provide relevant information about the molecular orbitals involved in the CT process and a quantitative energetic diagram is given. Hence, a Markus-Jordner model is used to quantitatively link the measured switching rate with the specific parameters of the induced CT process. This provides relevant quantitative information on the CT dynamics such as the electronic coupling, the free energy of reaction or the reorganization energy of the reacting elements inside the single molecule at the nanoscale.

**RESULTS AND DISCUSSION**

**Properties of the switching device.** Following the adsorption of the NiTPP molecules on the Si(100) surface, the STM topographies of the main adsorption site ($M_1$ in Fig. 1a) shows a spatial distribution of charge density with four bright lobes and a symmetry axis indicated by the white dotted line $A_1$. A detailed description of the $M_1$ conformation is provided in the electronic supplementary information (Fig. S1). Others adsorption sites have been observed in a non-negligible proportion but will not be discussed in this article. Imaging $M_1$ at voltages higher than -2.0 V (I > 100 pA) can induce the modification of the initial $M_1$ conformation into two new molecular conformations named $M_{2u}$ and $M_{2d}$ as depicted in Figs. 1b and 1c. The activation of the $M_1 \rightarrow M_{2d}$ or $M_1 \rightarrow M_{2u}$ conformational changes can be obtained during the STM scanning process or by applying a current pulse on the initial $M_1$ conformation. Once this transformation is completed (i.e. $M_1 \to M_{2u/2d}$), the adsorbed NiTPP molecule have two bistable states switching reversibly from the $M_{2u}$ to the $M_{2d}$ conformations via a tunnel electron excitation.

The STM topographies of the conformations $M_{1u}$ and $M_{1d}$ appears to have a mirror symmetry that gives a first indication of a geometrical chirality[24]. To trigger the molecular switching, several local electronic excitations on the NiTPP molecule have been tested. The positions $P_A$ and $P_B$ are the locations where the STM tip should be placed to have a good switching efficiency (Fig. 1b). These two positions can be studied successively by applying the excitation at a unique position due to the symmetry of the molecule (red dot in Fig. 1b). Indeed, when the excitation is initially applied at the position $P_A$ on the $M_{2d}$ conformation, the STM tip apex, which stays at a fixed position during the excitation process, will be subsequently located at the position $P_B$ of the $M_{2u}$ conformation after the $M_{2u} \leftrightarrow M_{2d}$ switching (Fig. 1c). The excitation procedure is then ran as follows: (i) The STM tip apex is placed at the upper part of the molecule (e.g. red dot in Fig. 1b at $P_A$). (ii) The feedback loop of the STM is switched off and the bias is adjusted to a value varying from -2.0 V to -2.5 V. (iii) during the excitation time, the tunnel current intensity is adjusted by changing the tip surface distance and recorded (Fig. S2a). Subsequently to this excitation process, the feedback loop is turned on again and the STM recovers its previous scanning parameters. To check that the switching processes are similar from one molecule to the other, we have also performed a set of measurements in which the initial STM tip positions are located at $P_A$ or $P_B$ on the $M_{2u}$ conformation without observing significant differences.

From the current traces (see ESI in Fig. S2), one can deduce the various switching time needed to induce the conformational changes of the molecule from the $M_{2d}$ to the $M_{2u}$ ($t_{M2u \to M2d}$) or from the $M_{2u}$ to the $M_{2d}$ ($t_{M2d \to M2u}$) conformations. The excitation process is repeated a large number of times and an example of the ensuing histogram of the measured durations is obtained (Fig. S2b). Its exponential decaying distribution shows that the switching process follow a binomial law from which an average time $\tau_{exc.}$ can be determined to estimate the average switching rate $1/\tau_{exc.}$[25,26]. Hence, an quantitative estimation of the switching efficiency (per tunnel electron) can be traduced by the value $Y = e/(I_{exc.} \times \tau_{exc.})$, where $I_{exc.}$ is the excitation current, $e$ the charge of the electron, and $I_{exc.} \times \tau_{exc.}$ represents the average number of electrons needed to induce a switch at a given excitation voltage.

The quantum yields to activate the molecular switch in both directions $M_{2d}$ to $M_{2u}$ (i.e. $Y_{M2d \rightarrow M2u}$) and $M_{2u}$ to $M_{2d}$ (i.e. $Y_{M2u \rightarrow M2d}$) as a function of the sample bias is thus measured quantitatively (Fig. 1d) and show that the switching efficiencies decreases when the sample bias increases (Figs. 1d and S2a). The variation of quantum yield for a given sample voltage ($V_s$ = -2.5 V) as a function of the excitation current indicate that the yield $Y_{M2d \rightarrow M2u}$ is constant over a range of current varying within almost one order of magnitude (Fig. 1e). This result tells us that the switching events arise from a one electron process, initial step of the CT dynamics[11].

Note that when the NiTPP molecule switches from the $M_{2d}$ to the $M_{2u}$ conformations or reversibly, the global position of the molecule does not move compared to the surface[7,10,27]. Here, only the side distribution of the charge density located inside the molecule varies before and after the switch which endorses the intramolecular character of the switching device[20,28].

In order to make a direct relation between the CT processes and the intramolecular switching, it is crucial to describe precisely the structural changes of the adsorbed NiTPP molecule. The $M_{2u}$ and the $M_{2d}$ molecular conformations can be obtained from the $M_1$ adsorption site (Fig. S3), by changing only two bonds thus having four remaining chemical bonds with the silicon surface (red arrows in Figs 2a and 2b). As one can see, the asymmetric shape of the STM pictures of the two switching conformation arise from the position of the pyrrole groups located in the front side of the molecule that has also lost one chemical bond with the silicon surface (green arrows in Figs. 2a and 2b). In fact, the pyrrole group located at the lower left side of the $M_{2d}$ conformation corresponds to the position of the bright protrusion observed in the STM topography (green arrow in the insert in Fig. 2a). The same effect is observed for the upper left pyrrole group of the $M_{2u}$ molecular conformation (green arrows in the insert of Fig. 2b). The two simulated conformations allow reproducing accurately the ensuing STM topographies observed experimentally (Figs 2c and 2d). In the side view of the conformations $M_{2d}$ and $M_{2u}$ (Figs. 2e and 2f) one can see now that each lateral aryl groups of the NiTPP molecule are detached from the surface, two being located above the back-bond row of the silicon surface providing to them a physisorbed characteristic. Looking at the front views in Figs. 2g and 2h, we can notice the mirror symmetry between the two switching conformations $M_{2u}$ and $M_{2d}$ implying that the two switching conformations are chiral[29,30].

**Relationship between switching and charge transfer.** From our experimental data (Fig. 1), one can deduced that the electronic excitation process responsible of the molecular switching at the points $P_A$ and $P_B$ in the range -2.5 V to -2.0 V involve the creation of a transient local cationic state. This process arise from the extraction of an electron from the adsorbed NiTPP molecule to the STM tip. More precisely, this implies the formation of a transient and local hole at the respective excited aryl groups (see the position A and B in Figs. 2a and 2g). Because the two corresponding aryl groups have no direct chemical bonds with the surface, the ensuing quantum cationic state can dynamically evolve with a relatively weak interaction with the Si(100) surface. Hence, the formation of a local cation followed by a CT process towards the porphyrin macrocycle is most probably at the origin of the observed switching events, since the experimental excitations can be fully related to the local properties of the NiTPP molecule.

To understand how a CT process can control the intramolecular switching device, we have measured the dI/dV signal at the two excitation locations $P_A$ and $P_B$ defined previously (Fig. 3a). The ensuing curves show two main DOS peaks at -2.5 V and -2.25 V when the signal is acquired on the aryl group B (black curve in Fig. 3a) located near the raised pyrrole group (Figs. 2g) whereas the dI/dV signal only increases at -2.5 V when the position A is probed (red curve in Fig. 3a). For comparison, the projected density of state (PDOS) is also computed over the aryl groups A and B and at the C-Si bond that links the front pyrrole group of the porphyrin macrocycle to the surface (see the blue arrow in Fig. 2g). The corresponding PDOS curves exhibit several comparable information with the dI/dV signals (Fig. 3b). Firstly, the PDOS curve computed over the aryl group at position B shows one main peak centered at ~ -2.5 eV with a second shoulder peak centered at -2.3 eV. Secondly, the PDOS curve integrated over the aryl group at position A (Fig. 3b) exhibits a single PDOS peak centered at -2.4 eV, similarly to the dI/dV signal in Fig. 3a. Both curves reveal that each excited aryl group located at A or B has two different electronic signatures. Interestingly, the PDOS curve at the C-Si bond (blue curve in Fig. 3b) does not show any signal in the previously cited energy window but a much less intense PDOS peak centered at ~ - 0.36 eV. A careful look at the data within this new energy range (see insert in Fig. 3b) indicate that the PDOS curve computed on A exhibits a peak at - 0.36 eV in relation with the C-Si bond while the PDOS curve on B shows a PDOS peak centered at - 0.03 eV in relation with the same chemical bond. Note that the computed PDOS peaks at the C-Si bond cannot be accurately measured experimentally since their energies are located in the band gap window of the bare Si(100) surface.

The switching quantum yields presented in Fig. 1d can be compared with the relative intensities of the dI/dV signals shown in Figs. 3a. When the excitation of the molecule is performed at -2.5 V (blue point N°1 in Figs. 3a), the quantum yield measured in B is lower than in A although the dI/dV intensities are similar. When the excitation bias increases to -2.2 V, the two yields $Y_{M2d \rightarrow M2u}$ and $Y_{M2u \rightarrow M2d}$ are still decreasing although the dI/dV signal increases at the excitation point B (blue point N° 2 in Fig. 3a at -2.25 V). Finally, an excitation at -2 V at the points A or B involves only very weak switching events as shown in Figs. 1d as the dI/dV signals are weak at this energy. Note that the same trend can be observed on the PDOS curves in Fig. 3b (blue points N° 1, 2 and 3). The intensity of the dI/dV signal is proportional to the local density of state, and therefore, the probability to create a local cation in the molecule has often been assumed to follow the same rule. However, our analysis shows that the measured switching efficiency is not merely related to the formation of a cation at a specific point inside the NiTPP as in this case, the evolution of the measured quantum yield would follow the variation of DOS intensity. Hence, the measured quantum yields are not a signature of the ionization probability of the (locally) excited NiTPP molecule but mainly traduce the efficiency of the quantum dynamics evolution of the local cation in relation to the molecular switching process. Therefore, the intramolecular dynamics can be quantitatively related to the measured switching yield and thus to the subsequent CT process dynamics.

To confirm this effect, we plot the cross-sectional local density of state (LDOS) distribution at various energy windows for a selected plane oriented parallel to the silicon surface that crosses the two lateral aryl groups selected for the electronic excitation and the two switching pyrroles groups at the front side of the molecule (blue line in Fig 2g). At -2.0 V, almost no LDOS is observed except at the central Ni atom (Fig. 3c). At -2.2 V (Fig. 3d) the LDOS is distributed over the entire molecule as an indication of a better electronic coupling between the aryl groups A or B and the porphyrin macrocycle. At an excitation bias of -2.5 V (Fig. 3e), the spatial distribution of the LDOS shows a high intensity located at the two positions A and B of the lateral aryl groups of the molecule. It is pooled with a weak hybridization of partial charge density distributed over the porphyrin macrocycle. These observations confirm that the switching process is mainly correlated to the electronic structure of the adsorbed NiTPP molecule with a resonant state located at the two lateral aryl groups A and B when the excitation occurs at -2.5 V. As previously mentioned, the DOS peak measured at -2.25 V in the dI/dV signal (Fig. 3a) in relation with the shoulder observed at -2.3 eV in the calculated PDOS curve (Fig. 3b) is traduced by a delocalization of the partial charge density between the lateral aryl groups A and B, and the porphyrin macrocycle. So far, it appears that this molecular state is not mostly involved in the observed switching process as it has no visible effects on the switching quantum yield measured experimentally.

The variations of the quantum yields can be related with the CT processes induced in the NiTPP molecule via the PDOS resonances located at the C-Si bond of the attached pyrrole group (Fig. 3b). Looking at the LDOS distributions calculated for the PDOS peaks centered at - 0.03 eV (Porph$^N$2 in Figs. 3f and 3h) and - 0.36 eV (Porph$^N$1 in Figs. 3g and 3h) one can see that the LDOS spreads on the left part of the porphyrin macrocycle with a clear delocalization on the lateral aryl group A for the PDOS peak centered at - 0.36 eV (red arrow in Fig. 3g). However, at the second PDOS peak centered at - 0.03 eV, the LDOS distribution is located at the right part of the NiTPP molecule with a fraction of charge density spreading inside the right lateral aryl group (black arrow in Fig. 3f).

This result clearly indicates that the excitation efficiencies of the two aryl groups A and B are correlated with two specific states having an electronic hybridization with the porphyrin macrocycle. Such electronic configuration is very favorable to enhance a CT process whose dynamics is specific to the excitation positions A or B. From these observations, it is possible to provide a description of the CT dynamics at the origin of the molecular switching. This model takes into account the evolution of a hole transfer triggered in one of the lateral aryl groups towards the porphyrin macrocycle. Based on the Antoniewicz approach that considers the presence of the silicon surface[11], Fig. 4a describes the case in which the aryl group located in A (Aryl A) is initially electronically excited via the tunnel electrons. The harmonic potential energy surface (PES) of Aryl A is represented in its ground state since the experiments are performed at 9 K. During the excitation, an electron from the Aryl A is extracted and tunnel to the STM tip to create a local transient cation which is described by the red PES (Aryl A$^+$). The excitation energy $E_{exc.}$ almost entirely corresponds to the selected bias $V_{exc.}$ ($E_{exc.} \sim qV_{exc.}$) due to only weak potential drops in the STM junction. Tuning $qV_{exc.}$ allows to reach various vibrational excited states in the cation PES curve. In these conditions, the vibrational energy levels of the PES curve of the Aryl A+ can match the ones in the PES of the state calculated at ~ - 0.36 eV (Porph.$^N$1) as shown in Fig. 4a. According to the Markus-Jordner model[31], the rate of a hole transfer from the aryl A$^+$ to the neutral porphyrin fragment is, in part, ruled by the matching of the vibrational states in both PES (black arrow 'CT' in Fig. 4a) and thus influence the hole transfer process via Franck-Condon factors similar to the one in a vertical transition. The electronic coupling parameter of the CT process is also important and related to the spatial hybridization of the coupled states. The LDOS distribution between the Aryl A$^+$ and the Porph.$^N$1 states are both showing spatial densities of state that overlap at similar energies. The matching of these parameters (vibrational and electronic coupling) can thus be different when the cation is created at position A compared to the one in B. When the hole is transferred to the porphyrin macrocycle, the Aryl $A^+$ is neutralized (Aryl $A^N$) and because of the presence of a local image charge in the silicon surface, the aryl $A^+$ returns to its neutral state with a slight additional kinetic energy $E_{ke}$ (Fig. 4b). The resulting cationic quantum state Porph. $^+1$ (red PES in Fig. 4b) then evolves rapidly over the whole porphyrin macrocycle PES to trigger the switching motion. A detailed description of the various molecular movements conducting to the switching is provided in the ESI document in Fig. S4. The subsequent neutralization of the porphyrin macrocycle is then likely to happen via the various chemical bonds that exist between the macrocycle and the surface. A similar quantum dynamics scheme can be proposed when the aryl B is electronically excited, one of the main difference in this case reside in the degree of hybridization between the aryl B group and the second porphyrin state which energy is computed at higher energy (- 0.03 eV) changing as well the corresponding Frank-Condon factors between these two states.

**Influence of the surface on the CT dynamics.** We can now discuss the influence of the interaction of the molecule with the surface on the CT dynamics as described in the previous section. Several processes that would be likely to trigger the molecular switch can be in competition: (i) When the excitation creates a local ion at the silicon surface, i.e. underneath the aryl group A or B, the ensuing ionic surface state might propagate through the surface and the molecular switch be triggered via substrate mediated vibrational excitations[32]. The chosen excitation positions at the two aryl groups A or B are located at the silicon back-bond row for which the two first occupied band energies usually lies between -1 to -1.8 eV and -3 to - 4 eV[33]. Hence, a surface mediated switching process can be ruled out since the excitation is optimized for the molecular resonant state at ~ -2.5 V, which does not match the one of the silicon back-bond surface states. In addition, a molecular switching induced via surface vibrational excitation would not provide significant differences of switching yield in relation with the excitation positions since the involved surface states are delocalized over the entire surface dimer and back-bond rows. In the same trend, no molecular switching events can be induced when the excitation is applied directly on the silicon surface, even very near the molecule, which further indicate that the observed excitation process involve molecular electronic states. (ii) The dynamic of the CT process can be influenced by the presence of the image charge in the silicon surface due to the presence of a transient cation in the NiTPP molecule[34]. In the present case, the image charge can favors the neutralization of the local excited Aryl $A^+$ (Fig. 4a) at the first stage of the cation relaxation dynamics, i.e. before than the CT occurs. Hence, the neutralization process can bring additional kinetic energy to the NiTPP molecule. Although the vibrational energy gained in this process might not be sufficient to perform the entire switching cycle of the porphyrin, this scenario would imply that there is no bias dependence on the measured quantum yield at the two chosen excitation locations since the gained kinetic energy depends only on the image charge potential. Hence, this scenario can also be ruled out as it is in contradiction with our experimental observations. (iii) The presence of the surface implies that the NiTPP molecule is chemically bonded at several locations on the silicon surface, situation that change the initial ro-vibrational spectrum of the molecule known in the gas phase. However, our results show that only one C-Si bond of the molecular switch is involved in the switching process. This further indicate that, although the surface affect the electronic and vibrational structure of the NiTPP molecule ground state due to its adsorption, the hole transfer mechanism is relatively efficient as it is related to a resonant electronic excitation that couples vibronically the excited cationic state of a lateral aryl group to the porphyrin macrocycle.

From this, although the silicon surface has a clear influence on the stationary electronic states of the adsorbed NiTPP before the excitation process, the CT dynamics induced via the STM excitation is likely to occur during the quantum evolution of the ionic state.

**Quantitative relationship between CT dynamics and switching yields.** Based on our findings, we can use the energetic diagram presented in Fig. 4 to estimate quantitative data in relation with CT dynamics via a simplified Markus-Jordner model. The hole transfer rate can thus be expressed as[35]:

$$k_{ET} \propto \frac{1}{\hbar} \sqrt{\frac{\pi}{E_r^0 k_B T}} |V|^2 exp\left\{-\frac{(E_r - \Delta E)^2}{4 E_r^0 k_B T}\right\} \quad (1)$$

Here, $\Delta E$ is the free energy of reaction between the two considered molecular elements (Fig. 4a), $V$ is the electronic coupling which is traduced by the spatial overlap between the initial excited cationic orbital of the aryl groups and the porphyrin macrocycle orbital which energies lies near the Fermi level of the surface. In view of our experimental conditions (9 K), we can consider that the involved vibrational energies $\hbar\omega_i$ are such that $\hbar\omega_i/k_B T \gg 1$ and thus describe the CT as a quantum process. Hence, the term $E_r$, which describes the intramolecular reorganization energy of the system can be decomposed into two terms $E_r = E_r^0 + \Delta E_r$. The first term $E_r^0$ is the remaining intramolecular reorganization energy and the second term can be associated to the effective energy used to bring the molecule into an initial distribution of vibrational levels such that $\Delta E_r \sim \gamma(qV_{exc.})$. Each of these vibrational channels being potentially involved in the CT process dynamics (Fig. 4a). Considering the very short times scale of the local ionization/neutralization of the molecule (i.e. <1 fs)[36] compared to the intramolecular hole transfer dynamics, we can assume that the measured switching yield traduce the intramolecular hole transfer process which rate $k_{ET}$ is such that $Y = (e/I_{exc.})k_{ET}$. It is therefore possible to use the expression (1) to numerically fit the experimental quantum yields variation presented in Fig. 3 and deduce relevant quantitative information about the induced hole transfer process. The fitting is done using a Levenberg-Marquardt algorithm by fixing initially the electronic coupling $V$ and deduce the other parameters $E_r^0, \gamma$ and $\Delta E$ to finally optimize $V$. The results are reported in the Table 1 for the two movements $M_{2d} \rightarrow M_{2u}$ and $M_{2u} \rightarrow M_{2d}$. The ensuing fitting curves are shown as dotted lines in Fig. 2f. The best fits are obtained in both cases for $V$ in the order of magnitude of $10^{-10}$ indicating that variation of the molecular switching efficiency at point A or point B is not mainly driven by the difference of electronic coupling between the two excited location and the porphyrin macrocycle. The fitted values of the free energy of reaction $\Delta E$ are worth $0.13 \pm 0.03$ eV for $M_{2d} \rightarrow M_{2u}$ and $0.15 \pm 0.03$ eV for $M_{2u} \rightarrow M_{2d}$. These values are consistent with the energy diagram presented in Fig. 6a in which the local ionization potential energies of the aryl groups A or B can thus be estimated ~ 2.1 eV[37,38].

If we now compare the fitted values of $E_r^0$ and $\gamma$, we can notice that $\gamma$ is higher for the $M_{2d} \rightarrow M_{2u}$ than for the $M_{2u} \rightarrow M_{2d}$ movements. This indicates that the vibrational coupling between the two molecular elements is significantly lower for the $M_{2u} \rightarrow M_{2d}$ switching movement. This is consistent with our theoretical findings which indicate that the energy of the neutral porphyrin state (Porph.$^N$2) considered when the excitation occurs at the position B is at a higher energy (- 0.03 eV) compared to the excitation at the position A (- 0.36 eV). This result also implies that the electronically induced process involves less efficient Frank-Condon factors between the excited Aryl group and the porphyrin macrocycle at B compared to A. This effect is balanced by a lower residual reorganization energy $E_r^0$ when the excitation occurs at the position B which is consistent with our energetic diagram (Fig. 4a).

From our investigations, it appears that the CT process that rule the molecular switching is favored compared to other surface mediated processes and that the efficiency of the molecular switch studied when excited at two precise locations is mainly governed by the vibrational matching between the cationic and two others resonant states related the porphyrin macrocycle. Therefore, tuning the excitation energy and the vibrational coupling in the molecule by adjusting the excitation bias and choosing the excitation location allow to prepare an initial quantum state that modulate the efficiency of the switching process and hence the ensuing CT dynamics. It is important to emphasis that the studied CT process could be generalized to either a hole transfer or an electron transfer. Applying this method to various molecular complexes having reversible functions will open new perspectives to study CT at the nanoscale. Our method can thus be applied to model structures like molecular dimers or more complex assemblies such as molecular layers electronically decoupled from the surface[39] that forms interfaces to mimic actives regions of OLEDs or solar cells devices.

**CONCLUSION**

In summary, we have shown that intramolecular switching devices can be exploited to probe and describe CT dynamics at the nanoscale with a low temperature STM. The study is performed on a single NiTPP molecule adsorbed on the bare Si(100) surface that shows two reversible states. Switching between these two states is triggered via an electronically induced excitation with the STM that locally forms a transient cation. The quantum evolution of this cationic state is at the origin of an intramolecular charge transfer process. Our method uses the experimental quantitative data of the intramolecular switching yield at various locations and excitation energies to describe the ensuing CT process dynamics and estimate fundamental related values. This is achieved by using a simplified Markus-Jordner model to numerically adjust the experimental switching yield variations. Quantitative estimations of the electronic coupling, the free energy of reaction or the reorganization energy of the reacting parts inside the single molecule at the nanoscale can thus be provided.

**Methods:**

Experiments are performed with a low temperature (9K) scanning tunneling microscope (STM) working in ultra-high vacuum (UHV). The surfaces are prepared from highly doped (n-type, As doped, $\rho = 5$ m$\Omega$.cm) Si(100) samples. The bare silicon surface is reconstructed in a c(4x2) structure as explained in several previous works via multiple annealing cycles[40]. To minimize surface defects, the base pressure in the UHV is kept under 4 x $10^{-11}$ torr during this process. Consecutively, the NiTPP molecules are evaporated on the silicon surface by heating a Knudsen cell at ~550 K (i.e. below the NiTPP dissociation temperature ~ 673 K) to ensure that they stay intact prior to their adsorption. During this process, the surface is kept at low temperature (12 K) via a liquid helium cooling to warrant a soft landing on the substrate and reduce irreversible surface-molecule interactions. In addition, a low evaporation rate is chosen to reduce molecule clusters on the surface. These parameters are adjusted with a quartz balance to obtain a homogenous coverage of ~ 0.1 ML. The dI/dV signals measurements are performed via a lock-in amplifier set with a modulation frequency of 837 Hz and an amplitude of 50 mV. The dI/dV curves are acquired several times on the same position and each series are averaged for a given set of parameters. Series of dI/dV curves are also acquired at various STM tip height to control possible energy shift on the observed peaks that may be due to the tip induced band bending or by electrostatic field effect. The shown data represent normalized curves (dI/dV/(I/V)) to shrink these effects.

Density-functional Theory (DFT) calculations are performed using the plane wave basis Vienna *ab initio* simulation package (VASP)[41,42], implementing the generalized gradient approximation (GGA) by Perdew, Burke and Ernzerhof (PBE)[43] and the projector augmented wave (PAW) potentials[44,45]. In the NiTPP/Si(100) system we describe the C-$1s^2$, N-$1s^2$, Ni-[Ar] and Si-[Ne] core electrons with the projector augmented wave (PAW) potentials. Using a cutoff kinetic energy of 400 eV and Γ-point (only one *k*-point due to the considerable dimensions of the system), we converged the total energy to a value below 1 meV/f.u. (f.u. = formula unit Si-fcc). The ionic optimizations are performed until all the forces are below 0.02 eV.Å$^{-1}$. The NiTPP/Si(100) system is modeled with a five layers thick Si(100)-(6x8) silicon slab, separated by 20 Å along the *z* direction normal to the surface (vacuum thickness). The Si atoms at the bottom layer are passivated with hydrogen, and this Si-H layer is kept fixed during atomic relaxation. Considering experimental STM images, the NiTPP molecule is positioned in specific sites of the slab above the topmost Si-layer and the whole system is allowed to relax. The constant-current mode STM images for the NiTPP/Si(100) system with bias voltages are computed. For this purpose, the bSKAN code is used[46,47]. This program implements the Tersoff-Hamman[48] approximation and thus assumes that the tunneling current is proportional to the local density of states (LDOS) at the surface at the position of the tip. bSKAN calculates the LDOS using the real-space single-electron wavefunction of the slab computed previously with VASP. Notice that each point in the space has an associated LDOS value for a given bias voltage. Thus, the constant-current STM images are the contour of constant LDOS of the surface within the vacuum above the surface atoms.

**Acknowledgements:**

HP and JL would like to thank the computation time provided by the Extreme Science and Engineering Discovery Environment (XSEDE) by National Science Foundation in USA with grant number OCI-1053575 and XSEDE award allocation number DMR110088. DR would like to thank the Fédération Lumière Matière (LUMAT) for their financial support (EaRTH project). All the authors wish to thank the CNRS for funding of the international program for scientific collaboration (PICS) THEBES.

**Figures captions**

**Figure 1:** (a) (42.3 x 42.3 Å²) STM topography ($V_s = -2.0$ V, $I_t = 50$ pA) of the $M_1$ conformation of the NiTPP adsorbed on the Si(100) surface. (b) and (c) (42.3 x 42.3 Å²) STM topography ($V_s = -2.0$ V, $I_t = 50$ pA) of the $M_{2d}$ and $M_{2u}$ switching conformations, respectively. The index 'd' and 'u' are used to indicate the position of the charge density protrusion located downwards (Fig. 2b) or upwards (Fig. 2e) in the STM images, respectively. The red dots indicate the positions of the STM tip during the excitation process. (d) Variation of the switching quantum yields per electron as a function of the sample bias for the $M_{2d} \rightarrow M_{2u}$ (black points) and the $M_{2u} \rightarrow M_{2d}$ (red points) switching movements. The dotted curves are numerically fitting curves obtained from the Markus-Jordner model. (e) Variation of the measured switching quantum yield per electron as a function of the excitation current for the movement $M_{2d} \rightarrow M_{2u}$ at a sample voltage of -2.5 V.

**Figure 2:** (a) and (b) ball and sticks structures of the simulated $M_{2d}$ and $M_{2u}$ molecular conformations after DFT optimization. The Si, C, N, Ni and H atoms color are the same as in Fig. S3. The inserts in (a) and (b) recall the corresponding STM topographies in which the green arrows

indicates the position of the bright protrusion. The red arrows indicate the positions of the remaining chemical bonds between the molecule and the surface. (c) and (d) calculated STM images (- 2.0 eV below $E_f$) of the $M_{2d}$ and $M_{2u}$ conformations after their optimization as presented in (a) and (b). (e) to (h) ball and sticks representations of the side ((e) and (f)) and front ((g) and (h)) views of the $M_{2d}$ and $M_{2u}$ molecular conformations, respectively. The green arrows in (g) and (h) indicate the positions of the lifted pyrrole group responsible of the bright protrusions in the STM topographies. The blue line in (g) located by two red arrows indicate the side view of the plane selected to plot the LDOS distribution shown in Fig. 3.

**Figure 3:** (a) Measured dI/dV curves at position A (red curve) and B (black curve) when the molecule is in the $M_{2d}$ conformation. The light blue points numbered from 1 to 3 recall the sample biases used for the excitation of the molecule (Fig. 2f). (b) PDOS curves calculated at positions A (red curve), B (black curve) and at the C-Si bond (blue curve) for an energy window (E-$E_f$) lying from -3.5 eV to 0 eV. (c) to (e) two dimensional cross sectional partial LDOS distributions (see Fig. 2g for the selected plane) for an energy linewidth of ± 0.05 eV at the three excited energies at -2.0 eV, -2.2 eV and -2.5 eV. The aryls A and B are circled in (e) to recall the excitation positions. (f) and (g) two dimensional (27 x 23 Å²) cross sectional partial LDOS distributions at - 0.03 eV and - 0.36 eV (± 0.05 eV) in relation to the PDOS peaks (Porph$^N$2 and Porph$^N$1) as observed in (b). (h) Band diagram of the Si(100) surface indicating the energies of the PDOS peaks related to the LDOS distributions compared to the conduction (CB) and valence (VB) bands.

**Figure 4:** (a) Energetic diagram of the CT process when the excitation is applied on the aryl (A) of the NiTPP molecule. The black, red and green curves represent the harmonic PES for the neutral (Aryl A), cationic (Aryl A+) states and the neutral porphyrin resonance state Porph. $^N$1 (- 0.36 eV below $E_f$), respectively. The excitation energy $E_{exc.}$, the ionization potential energy IP, the reorganization energy $E_r$ are indicated as well as the free energy of reaction ΔE. On the right side, a simplified band diagram of the silicon surface with the involved orbitals are represented. (b) Second step of the CT process describing the neutralization of the Aryl A$^N$ cation via a hole transfer (green PES) providing to the Aryl A an additional amount of energy $E_{ke}$ (black PES) due to the image charge effect. The dotted green and red PES curves describe, respectively, the neutral (Porph.$^N$1) and the cationic (Porph.$^+$1) states before and after the hole transfer. A possible neutralization of the NiTPP molecule is shown by a resonant electron transfer from the surface to the molecule on the surface band diagram. CB and VB stands for the conduction and valence bands, respectively.

**Table 1:** Fitting parameters $E_r^0$, $\gamma$, $\Delta E$ and $V$ used to numerically adjust the experimental quantum yield measurement presented in Fig. 2f via the Markus-Jordner equation for the two excitation movements $M_{2d} \rightarrow M_{2u}$ and $M_{2u} \rightarrow M_{2d}$.